\documentclass[twocolumn,showpacs,preprintnumbers,amsmath,amssymb,prl]{revtex4-1}

\usepackage{graphicx}
\usepackage{dcolumn}
\usepackage{bm}

\newcommand{\cN}{{\cal N}}

\newcommand{\be}{\begin{equation}}
\newcommand{\ee}{\end{equation}}
\newcommand{\bea}{\begin{eqnarray}}
\newcommand{\eea}{\end{eqnarray}}
\newcommand{\hk}{\hspace{0.1cm}}

\newcommand{\rk}{\right)}
\newcommand{\lk}{\left(}

\newcommand{\bx}{\boldsymbol{x}}
\newcommand{\by}{\boldsymbol{y}}
\newcommand{\bz}{\boldsymbol{z}}

\newcommand{\bk}{\boldsymbol{k}}
\newcommand{\bp}{\boldsymbol{p}}
\newcommand{\bq}{\boldsymbol{q}}

\newcommand{\vA}{\vec{A}}
\newcommand{\vx}{\vec{x}}
\newcommand{\vy}{\vec{y}}

\newcommand{\vk}{\vec{k}}

\newcommand{\vz}{\vec{z}}

\renewcommand{\vec}[1]{\mbox{\boldmath$#1$\unboldmath}}

\begin{document}

\title{Chiral symmetry breaking in Hamiltonian QCD in Coulomb gauge}

\author{M. Pak and H. Reinhardt}
\affiliation{Institut f\"ur Theoretische Physik,
Auf der Morgenstelle 14,
72076 T\"ubingen, Germany}
\date{\today}
%



\begin{abstract}
Spontaneous breaking of chiral symmetry is investigated in the Hamiltonian approach to QCD in Coulomb gauge.
The quark wave functional is determined by the variational principle using an ansatz which goes beyond
the commonly used BCS-type of wave functionals and includes the coupling of the quarks to the transversal
spatial gluons. Using the lattice gluon propagator as input it is shown that the low energy chiral properties of
the quarks, like the quark condensate and the constituent quark mass, are substantially increased by the coupling of the quarks
to the spatial gluons. Our results compare favourably with the phenomenological values.
\end{abstract}

\maketitle

\textit{Introduction.}---%
The infrared sector of QCD is characterized by two non-perturbative phenomena: confinement and spontaneous breaking
of chiral symmetry. For $N_f$ massless quark flavours the QCD Lagrangian is invariant under separate global flavour rotations of the left-
and right-handed quarks
\be
\label{1}
U_L (N_f) \times U_R (N_f) = U_V (N_f) \times U_A (N_f) \hk .
\ee
The vector $(V)$ and axial-vector $(A)$ groups rotate left- and right-handed quarks in the, respectively, same and
opposite way. In the
quantum theory, $U_A (N_f)$ is spontaneously broken by quark condensation $\langle \bar{q} q \rangle \neq 0$
resulting in the generation of a constituent quark mass and the occurence of $N^2_f$ Goldstone bosons. The latter
can be identified with the (light) pseudoscalar mesons. Chiral symmetry is a good starting point for the $N_f = 3$ light quark flavours $u, d, s$. 
The finite (current) quark masses explicitly break
the axial group $U_A (N_f)$ and induce finite masses for the pseudoscalar mesons.

Originally the dynamical breaking of chiral symmetry was studied within QCD-inspired models
which share with QCD the property of chiral symmetry, e.g., the Nambu-Jona-Lasinio model \cite{Nambu:1961tp, Ebert:1985kz, Ebert:1994mf, Klevansky:1992qe}
(which, however, is not
confining), or models with a confining two-body interaction \cite{Finger:1981gm, Le Yaouanc:1983iy, Adler:1984ri}. In recent
years there have been also attempts to study chiral symmetry breaking in continuum QCD either by means
of Dyson-Schwinger equations \cite{Aguilar:2010cn,Fischer:2006ub} or by the renormalization group flow equations \cite{Gies:2001nw}. 
In this letter we study the
spontaneous breaking of chiral symmetry within a variational approach, which is the extension to full QCD of the Hamiltonian approach to Yang-Mills theory in Coulomb gauge 
developed in recent years \cite{Feuchter:2004mk, Reinhardt:2004mm, Epple:2006hv, Reinhardt:2007wh}. 

Variational calculations in Coulomb gauge Yang-Mills theory were started in
Ref.~\cite{Schutte:1985sd} and later on taken up in Ref.~\cite{Szczepaniak:2001rg}. 
Our approach  \cite{Feuchter:2004mk, Reinhardt:2004mm, Epple:2006hv, Reinhardt:2007wh} differs from that of Refs.~\cite{Schutte:1985sd, Szczepaniak:2001rg} 
in the
ansatz for the vacuum wave functionals and, more importantly, in the full inclusion of the Faddeev-Popov determinant
and in the renormalization, for more details see Ref.~\cite{Greensite:2011pj}. Our
 approach has given a quite satisfactory description of the essential infrared
properties of Yang-Mills theory: a linearly rising static quark potential \cite{Epple:2006hv}, a gluon propagator in accord with Gribov's
formula and with lattice data \cite{Burgio:2008jr}, a perimeter law in the 't Hooft loop \cite{Reinhardt:2007wh} and a dielectric function of the Yang-Mills
vacuum \cite{Reinhardt:2008ek} embodying the phenomenological picture of the bag model. Similar infrared properties were also obtained
recently in a renormalization group flow equation approach to Hamiltonian Yang-Mills theory in Coulomb
gauge \cite{Leder:2010ji}. 

In the present letter we extend the variational approach in Coulomb gauge to full QCD.
We use an ansatz for the vacuum wave functional of the quark sector which goes beyond the BCS-type
wave functional used e.g. in  Refs.~\cite{Finger:1981gm, Adler:1984ri} and includes the coupling of the quarks to the transversal spatial gluons \cite{footnote}. We will
show that, using either the lattice results or the results of our variational approach for the gluon sector, this
coupling changes the chiral properties of the quarks, like the quark condensate and constitutent mass, by 20\,--\,60\% and brings these
quantities in the region of their phenomenological values.

\textit{Hamiltonian approach to QCD in Coulomb gauge.}---%
The part of the QCD Hamiltonian in Coulomb gauge which contains the quark fields $\psi (\vx)$, is given by \cite{Christ:1980ku}
\begin{align}
\label{Hamiltonian}
H = H_\textsc{d} + H_\textsc{qgc} + H_\textsc{c} \, ,
\end{align}
with
\begin{align}
H_\textsc{d} = {} - i &\int d^3 x \, \psi^\dagger (\vx) \left[ \vec{\alpha} \cdot \boldsymbol{\partial} + \beta m_0 \right] \psi (\vx) \hk ,  \\
\label{QGC}
H_\textsc{qgc}= {} - g &\int d^3 x \, \psi^\dagger (\vx) \boldsymbol{\alpha} \cdot \vA(\vx)  \psi (\vx) \hk , \\
H_\textsc{c} = {} \frac{g^2}{2} &\int d^3 x d^3 y \rho^a (\bx) \hat{F}^{ab} (\bx, \by) \rho^b (\by) \nonumber \\
+ \frac{g^2}{2} &\int d^3 x d^3 y \mathcal{J}^{-1} \rho^a_{\mathrm{dyn}}(\bx) \mathcal{J} \hat{F}^{ab}(\bx, \by) \rho^b(\by) \nonumber \\
\label{HC}
+\frac{g^2}{2} &\int d^3 x d^3 y \rho^a(\vx) \hat{F}^{ab} (\bx, \by) \rho_{\mathrm{dyn}}^b(\by) \hk .
\end{align}
Here $\vec{\alpha}$ and $\beta$ are the usual Dirac matrices, $m_0$ denotes the current quark mass, and
$\boldsymbol{A} (\vx) = \vA^a (\vx) T^a$ is the transversal spatial gauge field in the fundamental representation $T^a$ of the gauge group.
Furthermore,
\begin{align}
\label{3}
\rho^a (\vx) &= \psi^\dagger (\vx) T^a \psi (\vx) \\
\rho_{\mathrm{dyn}}^a (\vx) &= f^{abc} A^b_i (\vx) \Pi^c_i (\vx)
\end{align}
are the quark and gluon color charge densities, respectively, with $f^{abc}$ being the structure constants of the gauge group and $\Pi^a_k (\vx) = \delta / i \delta A^a_k (\vx)$
being the canonical momentum operator of the gauge field.
Finally, $\mathcal{J}(A)= \mbox{Det}(- \hat{D} \partial)$ is the Fadeev-Popov determinant and
\be
\label{4}
\hat{F}^{ab} (\vx, \vy) = \langle \vx \rvert \bigl[ (- \hat{D} \partial )^{- 1} \left(- \partial^2 \right) (- \hat{D} \partial )^{- 1} \bigr]^{ab} \rvert \vy \rangle
\ee
is the so-called Coulomb kernel where $\hat{D}^{ab}_k = \delta^{ab} \partial_k + g f^{acb} A^c_k$
is the covariant derivative in the adjoint representation.\\
\noindent The Coulomb term $H_{\textsc{c}}$, Eq.~(\ref{HC}), arises from the resolution of Gauss's law, so that
gauge invariance is fully taken into account. The Yang-Mills vacuum expectation value of the Coulomb kernel
\be
\label{5}
\langle \hat{F}^{ab} (\vx, \vy) \rangle_{\textsc{ym}} = \delta^{ab} F (\vx, \vy)
\ee
defines the static non-Abelian color Coulomb potential $F (\vx, \vy)$. Lattice measurements show that this quantity rises linearly
at large distances, however, with a coefficient $\sigma_\textsc{c}$ (referred to as Coulomb string tension)
which is about two to three times the value of the Wilsonian string tension $\sigma_\textsc{w}$ 
\cite{Greensite:2004ke, Langfeld:2004qs,Voigt:2008rr,Nakagawa:2011ar}. In momentum space
\begin{align}
\label{6}
g^2  F (k) \rightarrow \frac{8 \pi \sigma_\textsc{c}}{k^4} \; , \qquad k \rightarrow 0 \;  .
\end{align}
Due to asymptotic
freedom at small distances the static non-Abelian color Coulomb potential  $F (\vx, \vy)$ behaves essentially like an
ordinary Coulomb potential (up to anomalous dimensions).

\textit{Variational approach to the quark sector of QCD.}---%
We assume the QCD vacuum wave functional to be of the form $|\Phi \rangle  = |\phi\rangle \otimes |\textsc{YM}\rangle$
with a Gaussian type-of wave functional for the Yang-Mills sector \cite{Feuchter:2004mk, Reinhardt:2004mm}
\begin{align}
\label{YM-wave-functional}
 \langle A | \textsc{YM} \rangle = \mathcal{J}^{-\frac{1}{2}}(A) \exp \left(-\frac{1}{2} \int A^a_i (\vx) \omega (\vx, \vy) A^a_i (\vy) \right) . 
\end{align}
For the quark sector we use the following trial ansatz
\be
\label{7}
\Big| \phi \Big\rangle = {\cN} \exp \lk - \int \psi^{\dagger}_{+} (\vx) K (\vx, \vy) \psi_- (\vy) \rk \Big| 0 \Big\rangle \, ,
\ee
where $\psi_\pm (\vx)$ are the positive (negative) energy components of the quark field $\psi (\vx)$ and
$| 0 \rangle$ is the vacuum wave functional of the free quarks defined by
\be
\label{8}
\psi_+ (\vx) | 0 \rangle = 0 = \psi^{\dagger}_{-}(\bx) | 0 \rangle \, .
\ee
With arbitrary kernel $K$ the state $| \phi \rangle$ is the most general Slater determinant which is
not orthogonal to the vacuum $| 0 \rangle$. For the kernel $K$ we assume the form
\be
\label{9}
K (\vx, \vy) = K_0 (\vx, \vy) + \int d^3 z \, \bm{\mathcal{K}} (\vx, \vy; \vz) \cdot \vA (\vz) \, ,
\ee
where $K_0$ and $\bm{\mathcal{K}}$ are determined by minimizing the energy density of the
quarks. Finally, ${\cN}$ is a normalization constant, which is a functional of the kernel $K$, Eq.~(\ref{9}),
and thus of the gauge field. Through the vector kernel $\bm{\mathcal{K}}$, the wave functional (\ref{7}) contains
the coupling of the quarks to the gluons. If this coupling is neglected (by putting $\bm{\mathcal{K}} = 0$)
the wave functional $| \phi \rangle$, Eq.~(\ref{9}), is of the BCS-type used in Refs.~\cite{Finger:1981gm,Adler:1984ri}, for which 
the expectation value of the quark-gluon coupling term $H_{\textsc{qgc}}$, Eq.~(\ref{QGC}), vanishes. 

Our strategy is as follows: We first take the expectation value of the quark Hamiltonian H, Eq.~(\ref{Hamiltonian}),
in the fermionic state $| \phi \rangle$, Eq.~(\ref{7}), and subsequently in the Yang-Mills vacuum state (\ref{YM-wave-functional}).
The fermionic expectation value
$\langle H \rangle_\textsc{f} = \langle \phi | H | \phi \rangle$
can be taken exactly (by using Wick's theorem)
since the wave functional $| \phi \rangle$ is a Slater determinant. It can be expressed in terms of the static quark
propagator (in the presence of the gauge field)
\be
\label{10}
G (\vx, \vx') = \frac{1}{2} \langle \phi | \left[ \psi (\vx), \psi^\dagger (\vx') \right] | \phi \rangle   \, ,
\ee
which for the state (\ref{7}) is given in a matrix notation by
\begin{align}
\label{11}
G & = \Lambda_+ (1 + K K^\dagger)^{- 1} \Lambda_+  + \Lambda_- (1 + K^\dagger K)^{- 1} K^\dagger \Lambda_+ + \nonumber \\
&+ \Lambda_+ (1 + K K^\dagger)^{- 1} K \Lambda_-  - \Lambda_- (1 + K^\dagger K)^{- 1} \Lambda_- + \nonumber \\
&+ \frac{1}{2} (\Lambda_- - \Lambda_+)  \, ,
\end{align}
where $\Lambda_\pm$  are the projectors on positive and negative energy states $\psi_\pm = \Lambda_\pm \psi$.
Note that the quark propagator (and accordingly $\langle H \rangle_\textsc{f}$) is via the kernel $K$, Eq.~(\ref{9}),
 a non-local and non-linear
functional of the spatial gauge field $\vA$.

For the gluonic expectation value $\langle \ldots \rangle_{\textsc{YM}} = \langle \textsc{YM} | \ldots | \textsc{YM} \rangle$ of $\langle H \rangle_\textsc{f}$ in the Yang-Mills vacuum state (\ref{YM-wave-functional}), 
we will restrict ourselves to two (overlapping) loops. To
this order, in the gluonic expectation value of $\langle H \rangle_\textsc{f}$ we can use the replacement
\be
\label{13}
\langle \dots (1 + K^\dagger K)^{- 1} \dots  \rangle_{\textsc{ym}} \to \langle \dots  (1 + \langle K^\dagger K \rangle_{\textsc{ym}} )^{- 1} \dots \rangle_{\textsc{ym}} \,
\ee
and furthermore, replace the Coulomb kernel (\ref{4}) by its vacuum expectation value, i.e.\ by the static non-Abelian color Coulomb potential (\ref{5}).
It is then straightforward, although quite involved,
to evaluate $\langle \langle H \rangle_\textsc{f} \rangle_{\textsc{ym}}$ and take the variation with respect to the
kernels $K_0, \bm{\mathcal{K}}$ and $\omega$. In the present study we ignore the back-reaction of the gluon sector to the presence of the quarks, i.e.\ we keep the gluonic kernel (the gluon quasi-particle energy) $\omega$ at its form determined in the pure
Yang-Mills sector \cite{Feuchter:2004mk, Reinhardt:2004mm, Epple:2006hv}. We also ignore the small current quark mass $m_0$, which can be easily included if desired.
Furthermore, inspired by the form of the quark Hamiltonian $H$, Eq.~(\ref{Hamiltonian}), we assume for the quark kernels (\ref{9})
of the wave functional (\ref{7}) the form
\bea
\label{14}
K_0 (\vx, \vy) & = & \beta S (\vx - \vy) \nonumber\\
\bm{\mathcal{K}} (\vx, \vy; \vz) & = & \vec{\alpha} V \lk \vx - \vy, \vz - \vy \rk \, ,
\eea
where $S (\vx - \vy)$ and $V (\vx - \vy, \vz - \vy)$ are scalar functions. In principle, more complex
tensor structures of the quark-gluon coupling can be included as has been done in three-dimensional QED in
Landau gauge \cite{Kizilersu:2009kg}. 

\begin{figure}[t]
\centering
\includegraphics[width=70mm,angle=270]{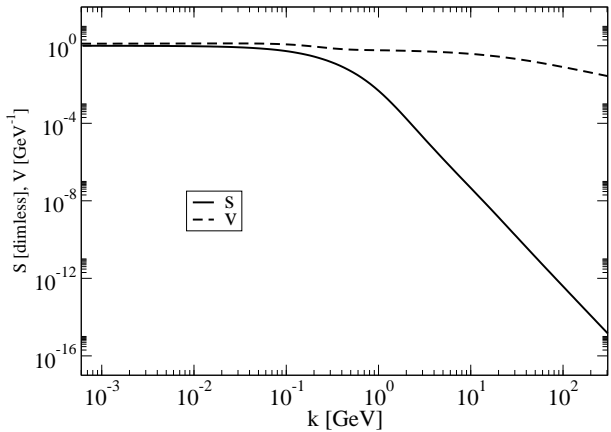}
\caption{\sl The variational kernels $S$ and $V$ for $\sigma_{\textsc{c}} = 2 \sigma_\textsc{w}$.}
\label{fig-s-v-asymptotic}
\end{figure}

Variation of the energy $\langle \langle H \rangle_{\textsc{f}} \rangle_{\textsc{ym}}$ with respect to the kernels Eq.~(\ref{14}) $S, V$ 
yields the following gap equations in momentum space 
\begin{align}
\label{15}
S (\bk) & =   \frac{I_\textsc{c}^{(1)}(\bk)}{|\bk| - I_{\textsc{qgc}}(\bk)}  \\
\label{16}
V (\bk,\bp) & = - \frac{g}{2} \, \frac{1+S^2(\bk)+R(\bk)}{|\bk| -I_{\textsc{qgc}}(\bk) + I_\textsc{c}^{(2)}(\bk)} \; ,
\end{align}
where
 \begin{align}
 R(\bk) =  C_F \, \int \frac{d^3 q}{(2\pi)^3} \, D(\vec{\ell}) \, V^2(\bk,\bq) 
\left[1+ (\hat{\bk} \cdot \hat{\vec{\ell}}) \, (\hat{\bq} \cdot \hat{\vec{\ell}})  \right] \, ,
 \end{align}
with $\vec{\ell}=\bk-\bq$, $C_F=(N_\textsc{c}^2-1)/(2 N_\textsc{c})$ being the quadratic Casimir and $D(\vec{\ell})$ being the static gluon propagator. In the above equations we have introduced the loop integrals
\begin{align}
I_{\textsc{qgc}}(\bk)  = {} &g C_F \,  \int \frac{d^3 q}{(2\pi)^3} \, D(\vec{\ell}) \, V(\bk,\bq) 
\left[ 1+ (\hat{\bk} \cdot \hat{\vec{\ell}}) \, (\hat{\bq} \cdot \hat{\vec{\ell}})  \right], \\
I^{(1)}_\textsc{c}(\bk) = {} &\frac{g^2}{2} C_F \int \frac{d^3 q}{(2\pi)^3} \frac{F(\bk-\bq)}{1+ S^2(\bq)+R(\bq)}
\nonumber \\ \times &\Big[ S(\bq) (1-S^2(\bk) + R(\bk)) + \nonumber \\ &- (\hat{\bk} \cdot \hat{\bq}) S(\bk) \left(1-S^2(\bq)
- R(\bq) \right)\Big]  , \\
I^{(2)}_\textsc{c}(\bk) =  {} &\frac{g^2}{2} C_F \int\frac{d^3 q}{(2\pi)^3}
\frac{F(\bk-\bq)}{1+S^2(\bq)+R(\bq)} \nonumber \\ \times &\Big[2 S(\bk) S(\bq) + \frac{1}{2}(\hat{\bk} \cdot \hat{\bq}) \left( 1-S^2(\bq)
- R(\bq) \right)\Big] .
\end{align}
A closer inspection of Eq.~(\ref{16}) shows that the vector kernel $V (\bk, \bp)\equiv V(\bk)$ does not depend on the 
momentum $\bp$ which is conjugate to the coordinate 
$\bz-\by$ in (\ref{14}).
After taking the gluonic expectation value of the quark propagator (\ref{11}) one finds the quark condensate
\be
\label{17}
\langle \bar{\psi} \psi \rangle = - \frac{2 N_C}{\pi^2} \int d q \,  q^2 \frac{S(q)}{1+S^2(q)+R(q)} \,
\ee
and the dynamical quark mass
\be
\label{18}
M (k) = k \, \frac{2 S (k)}{1 - S^2(k) - R(k)} \, .
\ee
\textit{Numerical Results.}---%
For the static gluon propagator $D (q) = 1/(2 \omega (q))$ we can take either the result of the variational calculation \cite{Feuchter:2004mk, Reinhardt:2004mm, Epple:2006hv} or the lattice result \cite{Burgio:2008jr}. Both propagators
are very similar to each other and can be well fitted by Gribov's formula \cite{Gribov:1977wm}
$\omega (k) = \sqrt{k^2 + M^4/k^2}$ with $M \approx 880$ MeV.

\begin{figure}[t]
\centering
\includegraphics[width=70mm,angle=270]{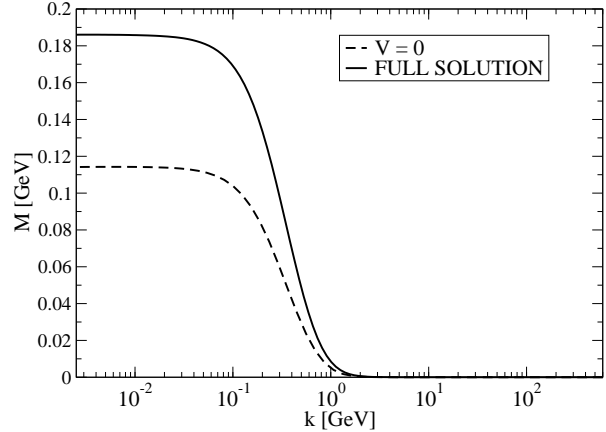}
\caption{\sl The dynamical mass $M$ for the full solution and for $V=0$ with $\sigma_{\textsc{c}} = 2 \sigma_\textsc{w}$.}
\label{fig-M}
\end{figure}

The quark Hamiltonian $H$, Eq.~(\ref{Hamiltonian}), like the full QCD Hamiltonian, is scale invariant.
In the present case the scale is set by the Coulomb string tension $\sigma_{\textsc{c}}$, which enters the static quark potential $F$, Eq.~(\ref{6}). 
Furthermore, the Hamiltonian $H$ contains the coupling
constant $g$ explicitly. In principle, after full renormalization the coupling constant should enter only in the form of the
running one. Since we are interested here in the infrared sector of QCD we will replace the running coupling
 calculated in the Hamiltonian approach in Ref.~\cite{Schleifenbaum:2006bq} by its (finite)
infrared limit. Furthermore, we use $\sigma_{\textsc{w}}= (440 \text{MeV})^2$ and $N_{\textsc{c}}=3$.
With this input the coupled equations (\ref{15}) and (\ref{16}) are solved numerically and the results are shown in Figs.~\ref{fig-s-v-asymptotic} 
and \ref{fig-M}. The quark condensate (\ref{17}) and the dynamical mass (\ref{18})
are affected in two ways by the inclusion of the coupling of the quarks to the spatial gluons: first through the change of
$S (\vk)$ and secondly through the occurence of the vector form factor $V(\vk)$ in the coupled equations (\ref{15}) and (\ref{16}). Neglecting the coupling
of the quarks to the gluons $(\bm{\mathcal{K}} = 0)$ we obtain for the quark condensate
\be
\label{21}
\langle \bar{\psi} \psi \rangle \approx \lk - 113 \, \text{MeV} \sqrt{\sigma_{\textsc{c}} / \sigma_{\textsc{w}}} \rk^3 \, ,
\ee
which agrees with the findings of Ref.~\cite{Adler:1984ri} if the same value for the string tension is used. With the inclusion of the quark-gluon coupling this value is shifted to
\be
\label{22}
\langle \bar{\psi} \psi \rangle \approx \lk - 135 \, \text{MeV} \sqrt{\sigma_{\textsc{c}} / \sigma_{\textsc{w}}} \rk^3 \, ,
\ee
which is a 20\% increase of the figure in the bracket. Lattice calculations show that $\sigma_\textsc{c} = (2\ldots3) \sigma_\textsc{w}$,
which yields for the quark condensate (\ref{22})
\be
\label{23}
\langle \bar{\psi} \psi \rangle \approx \lk - (191\ldots234) \, \text{MeV} \rk^3 \, ,
\ee
which compares favourably with the phenomenological value $\langle \bar{\psi} \psi \rangle = \lk - 230 \, \text{MeV} \rk^3$.
An even larger effect of the quark-gluon coupling is obtained for the constituent mass.
Neglecting the quark-gluon coupling ($\bm{\mathcal{K}}=0$) yields
\be
\label{25}
M \approx 84 \, \text{MeV} \sqrt{\sigma_{\textsc{c}} / \sigma_{\textsc{w}}} \, ,
\ee
while with this coupling included one finds
\be
\label{25}
M \approx 132 \, \text{MeV} \sqrt{\sigma_{\textsc{c}} / \sigma_{\textsc{w}}} \, ,
\ee
which is an increase of 57\%. Using again the values for $\sigma_\textsc{c}$ quoted above we obtain
\be
\label{27}
M \approx (186\ldots230) \, \text{MeV} \, .
\ee
\textit{Conclusions.}---%
Our results show that the coupling of the quarks to the spatial gluons, neglected in previous investigations \cite{Finger:1981gm, Le Yaouanc:1983iy,Adler:1984ri}, quite substantially influences the infrared chiral properties of the quarks. Typically the effect of the quark-gluon
coupling is in the range between 20\,--\,60\%. To make more accurate predictions within our approach, a more precise lattice
measurement of the Coulomb string tension $\sigma_\textsc{c}$ is required, which sets the scale in our approach.

\begin{acknowledgments}
\textit{Acknowledgments.}---%
The authors acknowledge useful discussions with G. Burgio, D. Campagnari, M. Leder, M. Quandt and P. Watson. This work was 
supported by the Deutsche Forschungsgemeinschaft (DFG) under contract Re856/6-3, by the 
Europ\" aisches Graduiertenkolleg ``Hadronen im Vakuum, Kernen und Sternen'' Basel-Graz-T\" ubingen and by the Graduiertenkolleg 
``Kepler-Kolleg: Particles, Fields and Messengers of the Universe''. 
\end{acknowledgments}

\end{document}